\documentstyle[12pt]{article}
\newcommand{\be}{\begin{equation}}
\newcommand{\ee}{\end{equation}}
\begin{document}
\begin{titlepage}
\vspace{1cm}

\begin{flushright}
CERN-TH/95-264\\
UWThPh-32-1995\\
hep-th/9510083
\end{flushright}
\vspace{1cm}
\begin{center}
{\Large \bf  Topologically Nontrivial Field Configurations
in  Noncommutative Geometry}\\[20pt]
{\small H. Grosse\footnote{Part of the Project P8916-PHY of the `Fonds
zur F\"orderung der wissentschaftlichen Forschung in \"Osterreich'.}  \\
Institut for Theoretical Physics, University of Vienna, \\
Boltzmanngasse 5, A-1090 Vienna, Austria \\[10pt]
C. Klim\v{c}\'{\i}k \\
Theory Division CERN \\
CH-1211 Geneva 23, Switzerland \\[10pt]
P. Pre\v{s}najder \\
Department of Theoretical Physics, Comenius University \\
Mlynsk\'{a} dolina, SK-84215 Bratislava, Slovakia\\[20pt] }
\begin{abstract}
In the framework of  noncommutative geometry we describe spinor
fields  with nonvanishing winding number on a truncated (fuzzy) sphere.
The corresponding field theory actions conserve all basic symmetries
of the standard commutative version (space isometries and global chiral
symmetry), but
due to the noncommutativity of the space the fields are regularized and
they contain only finite number of modes.
\end{abstract}
\end{center}
\vskip 0.5cm
\noindent CERN-TH/95-264\\
October 1995

\end{titlepage}

\section{Introduction}
The basic notions of the noncommutative geometry were developed in
\cite{Con1, Con2},
and, in the form of the matrix geometry, in \cite{Dub,DKM}. The essence of this
approach consists in reformulating first the geometry in terms of
commutative algebras and modules of smooth functions, and then generalizing
them to their noncommutative analogues.

In  standard field theory, to any point $x$ of some space(-time) manifold
$M$ the values of various fields are assigned:
\[
x \in M \longrightarrow \Phi (x)\ , A(x)\ ,...,
\]
as sections of some bundles over $M$, e.g. the line bundle of functions,
or the spinor bundle, etc. The smooth functions on $M$ form a commutative
algebra $\cal A = {\cal F}(M)$ with respect to the standard pointwise product:
$(fg)(x) = f(x)\ g(x),\ x \in M$. The bundle of smooth spinor fields
${\cal S}(M)$ on $M$ is an $\cal A$-module with respect to the
multiplication by smooth functions, which simply means that
any spinor can by multiplied by a scalar field. In the same way the linear
spaces of gauge and other
fields are $\cal A$-modules. If there exists
a sequence of deformations of the  commutative algebra $\cal A$ of the smooth
functions on the manifold $M$,  such that the deformed algebras are finite
dimensional, we may attempt to formulate a deformed field theory which would
possess just a finite number of degrees of freedom. Needless to say, this
may be a promising avenue towards a suitable regularization of
ill-defined field theoretical
path integrals.

Having the finite deformations of the algebra of scalar fields (= functions
on the manifold $M$) a further step has to be made in order to
obtain a physically useful regularization. It consists of building
up deformed spinor bundles, gauge fields, etc., which would be also just
the finite-dimensional vector spaces and at the same time modules of the
deformed algebra $\cal A$ of the scalar fields. For a fixed deformation
of $\cal A$ there may be many inequivalent deformations of the
corresponding modules. Therefore we would like to have a guiding principle that
would select the deformations which could legitimately be called  the
noncommutative spinor bundles or spaces of noncommutative connections.
In the general case of an arbitrary compact Riemannian manifold we are
still missing this principle; however, for  practical applications in
Euclidean field theories it is enough to understand the noncommutative
deformations
of the spheres $S^n$. Then the guiding principle reads: build up the
noncommutative modules, which would be representation spaces of the
symmetry group that rotates the sphere $S^n$.

 It is also natural  to  require
that the regularized field theory path integral respects the
rotational symmetry.
This  significantly narrows the room for the possible
deformations. Further restrictions  come from a claim that
some other desirable field theory symmetries are preserved in the deformed
level. Amazingly,  our experience so far shows that all relevant symmetries
can be incorporated while preserving the finiteness of the number of
degrees of freedom. To our knowledge this is an  unusual
(and  very favourable) property; for instance the fashionable lattice
regularization
does not enjoy it.

In  our previous
investigations \cite{GKP1,GKP2}, we  dealt with the two-dimensional
field theories, i.e.
the manifold
$M$ was the sphere $S^2$.   The real
scalar field on a truncated (= deformed) sphere was considered in \cite{GKP1}
and, recently,
we found a proper supersymmetric extension of this formalism \cite{GKP2}.
 In particular, the latter accounts for the description of the (chiral)
spinor fields
with vanishing topological charge  and gives a manifestly supersymmetric
regularization of two-dimensional supersymmetric theories. Historically,
the truncated two-sphere\footnote{Also referred to as `fuzzy',
`noncommutative' or `quantum' sphere in the literature \cite{Ber,Hop,Mad1}.}
was introduced in \cite{Ber}, where the deformed algebra of functions
emerged upon  geometric quantization of the (symplectic) volume form
on the sphere. Later the concept was rediscovered in \cite{Hop,Mad1,GP1}.
The first attempts to construct a field theory on the truncated sphere
were undertaken in \cite{Mad1,GM}; in \cite{GKP1,GKP2} we have
included the details of the perturbation expansion for
the deformed quantum scalar field, the construction of the deformed
chiral spinors and the regularization of the supersymmetric theories.

In this article, we continue these investigations by constructing the
deformed topologically nontrivial spinor bundles needed for the
inclusion of the monopoles. As  was argued in \cite{Y}, the correct
infinite volume limit (which means that the radius of the sphere approaches
infinity) requires the consideration of the monopole configurations
of the gauge fields interacting with the spinors. Hence, having in mind
our ultimate goal of the physical applicability of the construction, we
have to incorporate the spinor bundles with a nontrivial winding number.
Remarkably, this can be done already at the `kinematical' level of
the configuration space of the deformed field theory as opposed to the
case of the lattice regularization, where the topologically nontrivial
configurations emerge only dynamically \cite{Neu}.

 The plan of the paper is as follows: All basic notions we
need on complex scalar and spinor fields in the standard (commutative)
case are summarized in Section 2. In Section 3 we first describe the
topologically nontrivial configurations of a complex scalar field on a
standard sphere in a more algebraic language, and then we generalize them to
the noncommutative situation. In Section 4 we extend our approach to the
topologically nontrivial spinor field configurations on the noncommutative
sphere and write down a chirally symmetric field theory action.
 Section 5 contains concluding remarks.

\section{Topologically nontrivial fields on the sphere}
Here we  briefly describe the topologically nontrivial configurations of
complex scalar and spinor fields on the  standard two-sphere. The embedding
$S^2 \hookrightarrow {\bf R}^3$ is specified by the Cartesian coordinates
\be
x_1 \ =\ r \cos \varphi \sin \theta \ ,\
x_2 \ =\ r \sin \varphi \sin \theta \ ,\
x_3 \ =\ r \cos \theta \ ,
\ee
where $r>0$ is fixed,
$0\leq \theta \leq \pi$, $-\pi \leq \varphi \leq +\pi$.

The complex scalar field $\Phi$ on the upper hemisphere $V_+$ ($S^2$
without south pole $\theta \neq \pi$) is a function of the variables
\be
{\chi }'_1 = r^{1/2} \cos \frac{\theta}{2} \ ,\
{\chi }'_2 = -r^{1/2} \sin \frac{\theta}{2} e^{i\varphi } \ ,\
\ee
which are well defined on $V_+$:
\be
\Phi '\ =\ \Phi ' (\chi ' ,\chi '^* )\ =\
\sum a'_{m_1 m_2 n_1 n_2}
{\chi '}_1^{*m_1} {\chi '}_2^{*m_2} {\chi '}_1^{n_1} {\chi '}_2^{n_2} \ .
\ee
The monomials in this expansion are characterized by their phase on the
equator ($\theta =\pi /2$)
\[
e^{i(m_2 -n_2 )\varphi } \ =\
{\chi '}_1^{*m_1} {\chi '}_2^{*m_2}
{\chi '}_1^{n_1} {\chi '}_2^{n_2} |_{\theta =\pi /2} \ .
\]

In the same way, the complex scalar field $\Phi$  on the lower hemisphere
$V_-$ ($S^2$ without north pole $\theta \neq 0$) is a function of the variables
\be
{\chi ''}_1 = r^{1/2} \cos \frac{\theta}{2} e^{-i\varphi } \ ,\
{\chi ''}_2 = -r^{1/2} \sin \frac{\theta}{2} \ .\
\ee
Thus,
\be
\Phi ''\ =\ \Phi '' (\chi '' ,\chi ''^* )\ =\
\sum a''_{m_1 m_2 n_1 n_2}
{\chi ''}_1^{*m_1} {\chi ''}_2^{*m_2} {\chi ''}_1^{n_1} {\chi ''}_2^{n_2} \ .
\ee
Now, the monomials in this expansion have the following phase on the equator
($\theta =\pi /2$)
\[
e^{i(n_1 -m_1 )\varphi } \ =\
{\chi ''}_1^{*m_1} {\chi ''}_2^{*m_2}
{\chi ''}_1^{n_1} {\chi ''}_2^{n_2} |_{\theta =\pi /2} \ .
\]

As ${\tilde{\cal H}}_k , k \in \frac{1}{2} {\bf Z}$, we denote the line bundle
of sections  with the same expansion coefficients as in eqs. (3) and (5):
\[
\Phi '\ =\ \Phi (\chi ' ,\chi '^* )\ =\
\sum a_{m_1 m_2 n_1 n_2}
{\chi '}_1^{*m_1} {\chi '}_2^{*m_2} {\chi '}_1^{n_1} {\chi '}_2^{n_2} \ .
\]
\be
\Phi ''\ =\ \Phi (\chi '' ,\chi ''^* )\ =\
\sum a_{m_1 m_2 n_1 n_2}
{\chi ''}_1^{*m_1} {\chi ''}_2^{*m_2} {\chi ''}_1^{n_1} {\chi ''}_2^{n_2} \ ,
\ee
and with $k=\frac{1}{2}(m_1 + m_2 - n_1 - n_2 )$ fixed. On $V_+ \cap V_-$
they are related by the singular gauge transformation
\be
\Phi ' =e^{i\kappa \varphi}\Phi '' \ ,
\ee
where $\kappa = 2k$ is the so-called topological winding number.
Obviously, ${\cal A} = {\tilde{\cal H}}_0$ is an algebra, and
${\tilde{\cal H}}_k$ are ${\cal A}$-modules (with respect to the usual
multiplication by functions from ${\cal A}$).

The presence of the gauge transformation (7) requires the use of the covariant
derivatives:
\[
D'_{\mu} \ =\ i\partial '_{\mu} \ +\ A'_{\mu} \ ,\ \ {\rm on}\ V_+ \ ,
\]
\be
D''_{\mu} \ =\ i\partial ''_{\mu} \ +\ A''_{\mu} \, \ \ {\rm on}\ V_- \ .
\ee
Here we introduced the topological ($\kappa$-monopole) fields
\[
A'_{\mu} \ =\ i\kappa {\chi '}^+ \partial '_{\mu} \chi ' \ ,\ \
{\rm on}\ V_+ \ ,
\]
\be
A''_{\mu} \ =\ i\kappa {\chi ''}^+ \partial ''_{\mu} \chi '' \ ,\ \
{\rm on}\ V_- \ .
\ee
On $V_+ \cap V_-$ they are related by the singular gauge transformation (7):
\be
A'_{\mu} \ = A''_{\mu} \ -\ i h {\partial }_{\mu} h^{-1} \ ,\
h\ =\ e^{i\kappa \varphi } \ .
\ee

{\it Note}: We would like to stress that  the presence of the topological
$\kappa$-monopole field is dictated by the nontrivial topology, and not by the
dynamics of a system in question. The dynamical gauge field
${\cal A}_{\mu} = {\cal A}_{\mu} (x)$ is globally defined on $S^2$, and it
could be added to the topological one. In what follows, we shall not consider
this possibility.

The action for the complex scalar field on $S^2$ can be written in the form
\be
S[\Phi ,\Phi^* ]\ =\ \frac{1}{2\pi r} \int d^3 x\ \delta (x^2 - r^2)\
[\Phi^* D^2_{\mu} \Phi \ +\ V(\Phi^* \Phi )]\ ,
\ee
where $V(.)$ is a polynomial bounded from below.

In the same manner, the spinor field $\Psi$ we define
separately on the
upper and lower hemisphere:
\[
{\Psi }' \ =\ \left( \begin{array}{c}
{\Psi }'_1 \\
{\Psi }'_2 \end{array} \right) \ ,\ \ {\rm on}\ V_+ \ ,\
\]
\be
{\Psi }'' \ =\ \left( \begin{array}{c}
{\Psi }''_1 \\
{\Psi }''_2 \end{array} \right) \ ,\ \ {\rm on}\ V_+ \ .
\ee

As ${\tilde{\cal S}}_k$ we denote the bundle of spinor fields, which have their
components from ${\tilde{\cal H}}_k$:
\[
\Psi '_{\alpha} \ =\ \Psi_{\alpha} (\chi ' ,\chi '^* )\ =\
\sum a^{\alpha}_{m_1 m_2 n_1 n_2}
{\chi '}_1^{*m_1} {\chi '}_2^{*m_2} {\chi '}_1^{n_1} {\chi '}_2^{n_2} \ .
\]
\be
\Psi ''_{\alpha} \ =\ \Phi_{\alpha} (\chi '' ,\chi ''^* )\ =\
\sum a^{\alpha}_{m_1 m_2 n_1 n_2}
{\chi ''}_1^{*m_1} {\chi ''}_2^{*m_2} {\chi ''}_1^{n_1} {\chi ''}_2^{n_2} \ ,
\ee
where $k =\frac{1}{2}(m_1 + m_2 - n_1 - n_2 )$ is fixed. Obviously,
${\tilde{\cal S}}_k$ are ${\cal A}$-modules.

Alternatively, we can write the elements of the ${\tilde{\cal S}}_k$   in the
form
\[
{\Psi '} \ =\ f({\chi '},{\chi '}^* )\  \left( \begin{array}{c}
\chi '_1 \\
\chi '_2 \end{array} \right) \ +\
g({\chi '},{\chi '}^* )\  \left( \begin{array}{c}
{\chi '_2}^* \\
-{\chi '_1}^* \end{array} \right) \ ,\ {\rm on}\ V_+ \ ,\
\]
\be
{\Psi ''} \ =\ f({\chi ''},{\chi ''}^* )\  \left( \begin{array}{c}
{\chi ''}_1 \\
{\chi ''}_2 \end{array} \right) \ +\
g({\chi ''},{\chi ''}^* )\  \left( \begin{array}{c}
{\chi ''_2}^* \\
-{\chi ''_1}^* \end{array} \right) \ ,\ {\rm on}\ V_- \ .
\ee
The advantage of this form lies in the fact that both the $f$ and $g$ parts
of this decomposition are separately eigenfunctions of the chirality
operator
\be
\Gamma \ =\ \frac{1}{r} \sigma_i x_i \ ,\
\sigma_i {\rm -\ Pauli\ matrices}\ ,
\ee
with the eigenvalues $+1$ and $-1$ respectively:
\[
\Gamma {\Psi '} \ =\ f({\chi '},{\chi '}^* )\  \left( \begin{array}{c}
\chi '_1 \\
\chi '_2 \end{array} \right) \ -\
g({\chi '},{\chi '}^* )\  \left( \begin{array}{c}
{\chi '_2}^* \\
-{\chi '_1}^* \end{array} \right) \ ,
\]
\[
\Gamma {\Psi ''} \ =\ f({\chi ''},{\chi ''}^* )\  \left( \begin{array}{c}
\chi ''_1 \\
\chi ''_2 \end{array} \right) \ -\
g({\chi ''},{\chi ''}^* )\  \left( \begin{array}{c}
{\chi ''_2}^* \\
-{\chi ''_1}^* \end{array} \right) \ .
\]

The action of the (Dirac) spinor field can be written as
\be
S[\Psi ,\Psi^* ]\ =\ \frac{1}{2\pi r} \int d^3 x\ \delta (x^2 - r^2)\
[\bar\Psi D \Psi \ +\ W(\bar\Psi, \Psi )]\ ,
\ee
where $\bar\Psi =\Psi^+ \sigma^2$,
$W(.,.)$ is a gauge- and chiral-invariant potential describing the
selfinteraction of the spinor field, and $D$ is a Dirac operator defined
as
\[
D' \ =\ r\left[ i{\sigma '}^{\mu} (\partial '_{\mu} \ +\ A'_{\mu} )\
+\ \frac{1}{r} \right] ,\ \ {\rm on}\ V_+ \ ,
\]
\be
D'' \ =\ r\left[ i{\sigma ''}^{\mu} (\partial ''_{\mu} \ +\ A''_{\mu} )\
+\ \frac{1}{r} \right] , \ \ {\rm on}\ V_- \ ,
\ee
Here $\sigma^{\mu} = \sigma_i e^{\mu}_i$ ($e^{\mu}_i$ denote corresponding
zwei-beins). Only the topological ($\kappa$-monopole) gauge
field enters into $D$, and not a dynamical one.
The Dirac operator anticommutes with the
chirality operator,
\[
D\Gamma \ +\ \Gamma D\ =\ 0\ .
\]
This guarantees the chiral invariance of the action (16).

In the next two sections we first rewrite  all relevant formulas in more
algebraic terms, and then we introduce their truncated  noncommutative
analogues.

\section{Complex scalar field}
We start with  the Hopf fibration obtained from the mapping
${\bf C}^2 \rightarrow {\bf R}^3$ defined by
\[
\chi \ =\ \left( \begin{array}{c}
\chi_1 \\
\chi_2 \end{array} \right) \ \longrightarrow \ x\ =\ (x_1 ,x_2 ,x_3 )\ ,
\]
where
\be
x_i \ =\ \chi^+ \sigma_i \chi \ ,\ \ i=1,2,3\ .
\ee
The restriction $\chi^+ \chi = r > 0$ implies $x_i^2 = r^2$, i.e. we obtain
the fibration $S^3 \rightarrow S^2$ of a three-sphere in ${\bf C}^2$ with the
radius $\sqrt{r}$ onto the two-sphere in ${\bf R}^3$ with the radius $r$.
Since $x_i$ do not change under the transformation
\be
\chi \ \rightarrow e^{\frac{i}{2} \psi } \ \chi \ ,
\chi^+ \ \rightarrow e^{-\frac{i}{2} \psi } \ \chi^+ \ ,
\ee
we see that the fibre is $U(1)$.

As ${\cal H}_k , k \in \frac{1}{2}{\bf Z}$, we denote the linear space of
functions in ${\bf C}^2$ (or in $S^3$ after the restriction) of the form
\be
\Phi \ =\ \Phi (\chi ,\chi^* ) = \sum a_{m_1 m_2 n_1 n_2}
{\chi }_1^{*m_1} {\chi }_2^{*m_2} {\chi }_1^{n_1} {\chi }_2^{n_2} \ ,
\ee
with $k\ =\  \frac{1}{2} (m_1 +m_2 -n_1 -n_2)$ fixed (* denotes complex
conjugation). Under (19) the functions from ${\cal H}_k$ transform as
\[
\Phi \ \rightarrow e^{-ik\psi } \ \Phi \ .
\]
They are eigenfunctions of the operator
\be
K_0 = \frac{1}{2} [\chi^*_{\alpha} \partial_{\chi^*_{\alpha}} \ -\
\chi_{\alpha} \partial_{\chi_{\alpha}} ]
\ee
with  the eigenvalue $k$:
\[
K_0 \Phi = k \Phi ,\ \Phi \in {\cal H}_k \ .
\]
We have an involutive gradation
\[
{\cal H}^*_k \ =\ {\cal H}_{-k} \ ,\ {\cal H}_k {\cal H}_l
\subset {\cal H}_{k+l}
\]
with respect to the point-wise multiplication of functions
\be
(\Phi_1 ,\Phi_2)(\chi ,\chi^* ) =
\Phi_1 (\chi ,\chi^* )\ \Phi_2 (\chi ,\chi^* )\ .
\ee
The space ${\cal A} = {\cal H}_0$ endowed with the product (22) is a
commutative algebra, which is isomorphic to the algebra of all polynomials
in the variables $x_i , i =1,2,3$. Obviously, all ${\cal H}_k$ are
${\cal A}$-modules.

The differential operators
\be
J_k \ =\ \frac{i}{2} [\chi^*_{\beta} \sigma^{*k}_{\alpha \beta}
\partial_{\chi^*_{\alpha}}  - \chi_{\beta} \sigma^{k}_{\alpha \beta}
\partial_{\chi_{\alpha}} ]\ , \ k=1,2,3\ ,
\ee
map ${\cal H}_k$ to ${\cal H}_k$ and satisfy in ${\cal H}_k$ the
$su(2)$-algebra relations\footnote{It is important to note that the
operators $J_k$ and $K_0$ do commute with the restriction $\chi^+\chi=r$,
so they naturally act on the algebra of functions on $S^3$.}
\be
[J_i, J_j ] = i \varepsilon_{ijk} J_k \ .
\ee
The formulas
\[
J_j \ \chi_{\beta} \ =\ \frac{1}{2i} \sigma^j_{\alpha \beta}
\chi_{\alpha} \ ,\
J_j \ \chi^*_{\beta} \ =\ -\frac{1}{2i} \sigma^{*j}_{\alpha \beta}
\chi^*_{\alpha} \ ,
\]
guarantee that $\chi$ and $\chi^*$ transform like spinors under
transformations generated by (24), and consequently $x$ transforms like
a vector in ${\bf R}^3$. Moreover, the function
$C(x) = x^2_i = (\chi^+ \sigma_i \chi )^2$ satisfies
\be
J_i C(x)\ =\ 0\ ,\ i=1,2,3\ ,
\ee
i.e. $C(x)$ is an invariant function as expected.

Besides these operators we introduce operators $K_+$ and $K_-$ defined as
\be
K_+ \Phi \ =\ i \varepsilon_{\alpha \beta}\
\chi^*_{\alpha} (\partial_{\chi_{\beta}} \Phi )\ ,\
K_- \Phi \ =\ i \varepsilon_{\alpha \beta}\
(\partial_{\chi^*_{\alpha}} \Phi ) \chi_{\beta} \ .
\ee
They map ${\cal H}_k$ to ${\cal H}_{k+1}$ and
${\cal H}_{k-1}$ respectively. The operators  $K_{\pm}$ and $K_0$
satisfy $su(2)$ algebra relations
\[
[K_0 , K_{\pm} ]\ =\ \pm K_{\pm} \ ,\ [K_+ ,K_- ]\ =\ 2K_0 \ .
\]
 Only products $K_{\pm} K_{\mp}$ act in ${\cal H}_k$, and $K_0$ takes
there the
constant value $k$. The operators $K_0 ,\ K_{\pm}$ commute with $J_i$,
$i\ =\ 1,2,3$, but they are not independent as the corresponding Casimir
operators are equal:
\be
J_i^2 \ =\ K_0^2 \ +\ \frac{1}{2} (K_+ K_- \ +\ K_- K_+ )\ .
\ee
To any $\Phi \in {\cal A}$ we assign the standard integral over $S^2$
\be
I_{\infty} [\Phi ]\ =\
\frac{1}{2\pi r} \int d^3 x\ \delta (x^2_i -r^2 )\ \Phi (x)\ .
\ee
This allows the  introduction of  the scalar product on ${\cal H}_k$ as
follows:
\be
(\Phi_1 \Phi_2 )_k \ =\ I_{\infty} [\Phi^*_1 \Phi_2 ] \ .
\ee

We identify the
complex scalar field $\Phi$ with the topological charge $\kappa$
\footnote{In other words: the section of the line-bundle with the
winding number $\kappa$.}
 with the elements of ${\cal H}_k$. The corresponding field action
is given as
\be
S[\Phi ,\Phi^* ]\ =\ I_{\infty}
\Big[\frac{1}{2}\Phi^* (K_+ K_- + K_- K_+)\Phi \
+\ V(\Phi^* \Phi )\Big]\ ,
\ee
where $V(.)$ is a polynomial bounded from below. According to eq. (27) the
differential operator $\frac{1}{2}(K_+ K_- + K_- K_+ )$ can be rewritten in
${\cal H}_k$ as follows:
\be
\frac{1}{2}(K_+ K_- + K_- K_+ )\ =\ J^2_i \ -\ k^2 \ .
\ee
We stress that formula (30) for the action is equivalent to (11).

We obtain the noncommutative (fuzzy) line-bundles by replacing the
commuting parameters $\chi_{\alpha} ,\chi^*_{\alpha}$, $\alpha = 1,2$, by
the noncommutative ones, expressing them in terms of  annihilation and
creation operators as
\be
\hat\chi_{\alpha} \ =\ A_{\alpha} \ R^{-1/2} \ ,\
\hat\chi^*_{\alpha} \ =\ R^{-1/2} A^*_{\alpha} \ ,
\ee
where
\be
R\ =\ A^*_{\alpha} A_{\alpha} \ ,
\ee
so that the condition $\chi^*_{\alpha} \chi_{\alpha} = 1$ is satisfied
(without lack of generality, we
choose the unit radius  $r=1$  of the sphere).
The operators $\chi_{\alpha}$ are well defined on all vectors
except vacuum; we complete the definition by postulating that they
annihilate the vacuum.  The operators
$A_{\alpha}$ and $A^*_{\alpha}$ (* denotes Hermitian conjugation) act in the
Fock space ${\cal F}$ spanned by the orthonormal vectors
\[
|n_1 ,n_2 \rangle =
\frac{1}{\sqrt{n_1 ! n_2 !}} (A^*_1)^{n_1} (A^*_2)^{n_2} |0 \rangle \ ,
\]
where $|0 \rangle$ is the vacuum defined by
$A_1 |0 \rangle = A_2 |0 \rangle = 0$. They satisfy in ${\cal F}$ the
commutation relations
\be
[A_{\alpha} , A_{\beta} ]\ =\ [A^*_{\alpha} , A^*_{\beta} ]\ =\ 0\ , \
[A_{\alpha} , A^*_{\beta} ]\ =\ \delta_{\alpha \beta} \ .
\ee

The operators $R$ and
\be
R_j \ =\ \frac{1}{2}\ A^*_{\alpha} \sigma^j_{\alpha \beta} A_{\beta}
\ee
satisfy in ${\cal F}$ the $u(2)$ algebra commutation relations
\be
[R_i, R_j ] = i \varepsilon_{ijk} R_k \ ,\ [R_i, R] = 0\ .
\ee
Equation (35) is the Schwinger-Jordan realization of the $su(2)$ algebra.
On the other hand, it is just the noncommutative (quantum) version of the
Hopf fibration (18).

As ${\hat{\cal H}}_k , k \in \frac{1}{2} {\bf Z}$, we denote the linear space
spanned by the normal products
\be
\hat\chi^{*m_1}_1 \hat\chi^{*m_2}_2 \hat\chi^{n_1}_1 \hat\chi^{n_2}_2
\ee
with $k\ =\  \frac{1}{2} (m_1 +m_2 -n_1 -n_2)$ fixed. Obviously,
$\hat{\cal A} = {\hat{\cal H}}_0$ is the noncommutative algebra generated by
$R, R_j , j=1,2,3$  with   relations (36). The spaces ${\hat{\cal H}}_k$
are ${\hat{\cal A}}$-bimodules. The operators $J_j$ act in ${\hat{\cal H}}_k$
as follows
\be
J_j f\ =\ [R_j ,f]\ ,
\ee
and they satisfy in ${\hat{\cal H}}_k$ the $su(2)$ algebra commutations
relations.

For the following discussion, it is useful to consider $(N+1)$-dimensional
subspaces
\[
{\cal F}_N \ =\ \{ |n_1 ,n_2 \rangle \ ,\ n_1 +n_2 \ =\ N\ \}\ ,\
N\ =\ 0,1,2,...\ ,
\]
of the Fock space $\cal F$. The operator $R$ takes in ${\cal F}_N$ the
constant value $R=N$. The subspace ${\cal F}_N$ is the representation
space of the unitary irreducible spin $\frac{N}{2}$-representation of the
$su(2)$ algebra in which the Casimir operator
\be
C = R^2_3 \ +\ \frac{1}{2} (R_+ R_- \ +\ R_- R_+ )\ ,\
R_{\pm} \ =\ R_1 \pm i R_2 \ ,
\ee
takes the value
\be
C = \frac{N}{2} \left( \frac{N}{2} + 1 \right)  \ .
\ee

As ${\hat{\cal H}}_{MN}$ we denote the space of linear mappings from
${\cal F}_N$ to ${\cal F}_M$ spanned by the monomials (37) with
$m_1 +m_2 \leq M$, $n_1 +n_2 \leq N$, $m_1 +m_2 -n_1 -n_2 = M-N$.
Obviously,
\[
{\hat{\cal H}}^*_{MN} \ =\ {\hat{\cal H}}_{NM} \ ,\
{\hat{\cal H}}_{LM} {\hat{\cal H}}_{MN} \subset {\hat{\cal H}}_{LN} \ .
\]
Any operator $\Phi \in {\hat{\cal H}}_{MN}$ maps ${\cal F}_N$ to ${\cal F}_M$,
and can be represented by an $(M+1) \times (N+1)$ matrix. There is an
antilinear isomorphism between ${\hat{\cal H}}_{MN}$ and
${\hat{\cal H}}_{NM}$ represented as the (matrix) Hermitian conjugation.
In ${\hat{\cal H}}_{MN}$ we introduce the scalar product
\[
(\Phi_1 ,\Phi_2 )_{MN} \ =\ \frac{1}{J+1}\ Tr_N (\Phi^*_1 \Phi_2 ) \
\]
\be
=\ \frac{1}{J+1}\ Tr_M (\Phi_2 \Phi^*_1 )\ =\ (\Phi^*_2 ,\Phi^*_1 )_{NM} \ ,
\ee
where $J=\frac{1}{2}(M+N)$ and $Tr_{N'}$ denotes the trace in the space
${\hat{\cal H}}_{N'N'}$.

In particular, ${\hat{\cal A}}_N = {\hat{\cal H}}_{NN}$ is an
$(N+1)^2$-dimensional algebra generated by $R^{(N)}, R^{(N)}_j$, $j=1,2,3$,
where $R^{(N)}, R^{(N)}_j$ denote the restriction of $R$ and $R_j$ in
${\cal F}_N$. This restriction generates the algebra homomorphism
${\hat{\cal A}} \to {\hat{\cal A}}_N$. We point out that in ${\hat{\cal A}}_N$
there is an additional relation
\be
R^{(N)} - N\ =\ 0\, ,
\ee
which expresses the fact that
${\cal F}_N$ is the space of an irreducible representation of $su(2)$.
 To any operator $\Phi \in {\cal A}_N$ we
assign the integral
\be
I_N [\Phi ]\ =\ \frac{1}{N+1} Tr_N (\Phi )\ .
\ee
In \cite{GKP2}, we proved that
for $N\to\infty$ the algebras $\hat{\cal A}_N$
 approach the standard commutative algebra of functions ${\cal A}_{\infty}$ and
 $I_N [\Phi ] \to I_{\infty} [\Phi ]$.
Obviously, ${\hat{\cal H}}_{MN}$ is a left ${\hat{\cal A}}_M$-module and a
right ${\hat{\cal A}}_N$-module.

The generators of $su(2)$ rotations  $J_j$  in ${\hat{\cal H}}_{MN}$ are
given by
\be
J_j \Phi \ =\ R^{(M)}_j \Phi \ -\ \Phi R^{(N)}_j \ .
\ee
This $su(2)$ algebra representation is reducible and is equivalent to the
direct product of two irreducible $su(2)$ representations:
\be
\frac{M}{2} \otimes \frac{N}{2} \ =\
|k| \ \oplus \ (|k|+1)\ ...\ \oplus \ J\ ,
\ee
where $k=\frac{1}{2}(M-N)$ and $J=\frac{1}{2}(M+N)$. This means that any
operator $\Phi \in {\hat{\cal H}}_{MN}$ can be expanded into operators
$\Phi^{j}_{Jkm}$, belonging to the representations indicated in (45):
\[
J^2_i \Phi^{j}_{Jkm} \ =\ j(j+1) \Phi^{j}_{Jkm} \ ,\
j = |k|, |k|+1,\ \dots \ ,J\ ,
\]
\[
J_3 \Phi^{j}_{Jkm} \ =\ m \Phi^{j}_{Jkm} \ ,\ |m|\leq j\ .
\]
Putting $J_{\pm} = J_1 \pm iJ_2$ we obtain  the highest-weight
functions
\be
\Phi^{j}_{Jkj} \ =\ N_{Jkj} \sqrt{\frac{(2j+1)!}{(j+k)!(j-k)!}}\
\hat\chi_2^{*j+k} \hat\chi_1^{j-k} \ ,
\ee
satisfying $J_+ \Phi^{j}_{Jkj} = 0$. Here the normalization constant $N_{Jkj}$
 with respect to the scalar product (41) is given by the equation
\be N_{Jkj} \ =\ \sqrt{\frac{(J+1)(J+k)!(J-k)!}{(J+j+1)!(J-j)!}}\ee
(we used eqs. (26) on p. 608 and (43) on p. 618 of  Ref. \cite{Pru}).
The other normalized functions $\Phi^{j}_{Jkm}$, $m = 0,\pm 1,\dots ,\pm j$,
in the irreducible representation containing $\Phi^{j}_{Jkj}$ are given by
\be
\Phi^{j}_{Jkm} \ =\
\sqrt{\frac{(j+m)!}{(j-m)!(2j)!}} J^{j-m}_- \Phi^{j}_{Jkj}\ .
\ee

Now we are ready to discuss  the  commutative limit $J\to\infty$, $k$ fixed.
Not only in this
limit do
$\chi*$ and $\chi$ commute among themselves, also the normalization
factor $N_{Jkj}$ approaches $1$  and expression (46) becomes
 the standard Wigner
$D$-function $D^j_{kj}$, expressed in terms of
$\chi_{\alpha} ,\chi^*_{\alpha}$ instead of Euler angles. Since $[J_i ,R]=0$,
the same remains true for the functions $\Phi^{j}_{Jkm}$ given in (48).
The
normalization coefficient $N_{Jkj}$ is also a cut-off factor, as can be seen
from (47),
because $N_{Jkj}=0$ for $j>J$.
If we vary $k$ while keeping $J$ fixed, then   $\chi*$ and $\chi$
will cease to commute    for               $J-\vert k\vert\to 0$,
even though $J$ can be very large. This is in accordance with the general
principle
(cf. \cite{GKP2})
that
 approaching the maximal spin $J$ of the truncation the multiplication
becomes noncommutative.

In the noncommutative case we identify a section $\Phi$ of a complex line
bundle with fixed winding number with an  element of ${\hat{\cal H}}_{MN}$.
The corresponding field theory action we take in the form
\be
S_{MN} [\Phi ,\Phi^* ]\ =\ \frac{1}{J+1}\
Tr_N \Big[\frac{1}{2}\Phi^* (K_+ K_- + K_- K_+ )
\Phi \ +\ V(\Phi^* \Phi )\Big]\ ,
\ee
where in the noncommutative case the operators $K_{\pm}$ are defined by
\be
K_+ \Phi \ =\ i\varepsilon_{\alpha \beta}
A^*_{\beta} [\Phi ,A^*_{\alpha} ]\ ,\
K_- \Phi \ =\ i\varepsilon_{\alpha \beta}
[A_{\alpha} ,\Phi ]A_{\beta} \ .
\ee
Note that the `topological charge' operator $K_0$ defined as
\be
K_0 \Phi \ =\ \frac{1}{2} [R ,\Phi ]
\ee
takes in ${\hat{\cal H}}_{MN}$ the constant value $k=\frac{1}{2} (M-N)$.
 The order of operators in (50) is essential because it guarantees that the
operators $K_{\pm}$ act on monomials exactly in the
same way as in the commutative case.

{\it Note}: We would like to stress that for the description of topologically
nontrivial field configurations (with $\kappa \neq 0$), two algebras
${\hat{\cal A}}_M$ and ${\hat{\cal A}}_N$ (with $M-N=\kappa \neq 0$) are
needed. This is the reason why the discussion within  only one matrix algebra
($M=N$) corresponds to the topologically trivial situation (see e.g.
\cite{DKM,GKP1,GKP2,GM}).

If the winding number of the field $\Phi$ is not fixed, we work with
fields from the space
\be
{\hat{\cal H}}_{(J)} \ =\ \bigoplus_{M+N=2J} {\hat{\cal H}}_{MN} \ ,
\ee
and the corresponding action we take as
\be
S_{(J)} [\Phi ,\Phi^* ]\ =\ \sum_{M+N=2J} S_{MN} [\Phi ,\Phi^* ]\ .
\ee
The action (53) has the following basic properties:

1) it has the full $su(2)$ symmetry corresponding to the  rotations of $S^2$;

2) it describes a model with a finite number of modes since, in fact, it
corresponds to a particular matrix model; and

3) it approaches in the limit $J \to \infty$ the commutative action (for
any given polynomial field $\Phi$).

In general, the complex scalar field from ${\hat{\cal H}}_{(J)}$ can be
expanded as
\be
\Phi \ =\ \sum_{j=0}^J \sum_{k,m=-j}^{+j} \ a^j_{km} \ \Phi^{j}_{Jkm} \ .
\ee
The quantum field mean value of a functional $F[\Phi ,\Phi^* ]$ is
defined as
\be
\langle F[\Phi ,\Phi^* ] \rangle \ =\ \frac{\int D\Phi D\Phi^*
e^{-S_{(J)} [\Phi ,\Phi^* ]} F[\Phi ,\Phi^* ]}{\int D\Phi D\Phi^*
e^{-S_{(J)} [\Phi ,\Phi^* ]}} \ ,
\ee
where $D\Phi D\Phi^* \ =\ \prod_{jkm} da^j_{km} da^{*j}_{km}$ is a finite
product of the standard measures in the complex plane. The quantum mean
values are
well defined for any polynomial functional $F[\Phi ,\Phi^* ]$.

Under rotations specified by the Euler angles $\alpha ,\beta ,\gamma$ the
coefficients of the field expansion transform as
\[
a^j_{km} \to {a'}^j_{km'} \ =\ \sum_m
D^j_{m'm} (\alpha ,\beta ,\gamma)\ a^j_{km} \ ,
\]
\[
a^{*j}_{km} \to {a'}^{*j}_{km'} \ =\ \sum_m
D^{*j}_{m'm} (\alpha ,\beta ,\gamma)\ a^{*j}_{km} \ .
\]
These are the  unitary transformations not changing the
measure $D\Phi D\Phi^*$.
This completes the proof of rotational invariance of the model at the
quantum level.

\section{Spinor fields}
For construction of the topologically nontrivial  spinor fields
 we use the superspace approach developed in \cite{GKP2}.
 First we perform the ${\cal N}=1$ superextension of the Hopf
fibration described in the previous section.  We
obtain it from  the mapping ${\bf C}^{2,1} \rightarrow {\bf R}^{3,2}$ given by
\[
\xi \ =\ \left( \begin{array}{c}
\chi_1 \\
\chi_2 \\
a \end{array} \right) \ \to \ (x_i , \theta_{\mu} )\ ,
\ i=1,2,3\ ,\ \mu = +,-\ ,
\]
where
\be
x_i \ =\ \xi^+ \Sigma_i \xi \ ,\
\theta_{\mu} \ =\ \xi^+ F_{\mu} \xi \ .
\ee
Here $\xi^+ = (\chi^*_1 ,\chi^*_2 ,a^* )$, and
$\Sigma_{\pm} = \Sigma_1 \pm i\Sigma_2$, $\Sigma_3$ and
$F_{\pm}$ are $3 \times 3$ matrices
given by
\[
\Sigma_+ \ =\ \left( \begin{array}{ccc}
0 & 1 & 0 \\
0 & 0 & 0 \\
0 & 0 & 0 \end{array} \right) ,\
\Sigma_- \ =\ \left( \begin{array}{ccc}
0 & 0 & 0 \\
1 & 0 & 0 \\
0 & 0 & 0 \end{array} \right) ,\
\Sigma_3 \ =\ \frac{1}{2} \left( \begin{array}{ccc}
1 & 0 & 0 \\
0 & -1 & 0 \\
0 & 0 & 0 \end{array} \right) ,\
\]
\be
F_+ \ =\ \frac{1}{\sqrt{2}} \left( \begin{array}{ccc}
0 & 0 & -1 \\
0 & 0 & 0 \\
0 & -1 & 0 \end{array} \right) ,\
F_- \ =\ \frac{1}{\sqrt{2}} \left( \begin{array}{ccc}
0 & 0 & 0 \\
0 & 0 & -1 \\
1 & 0 & 0 \end{array} \right) .
\ee
The quantities
 $\chi_{\alpha}$, $\chi^*_{\alpha}\ $ $(\alpha = 1,2)$ and $a$, $a^*$
are respectively even and odd coordinates of the superspace $C^{2,1}$;
* denotes the graded involution \cite{GI,GKP2} characterized by the
properties
\[ (ab)^*=(-1)^{deg~a~deg~b}b^*a^*, \qquad a^{**}=(-1)^{deg~a}a.\]
Note that a restriction $\chi_1^*\chi_1 +\chi_2^*\chi_2 +a^*a=2\rho$ implies
$x_i^2+\theta_+\theta_-=\rho^2$. Thus we obtain the Hopf superfibration
$sS^3\to sS^2$ of the 3-supersphere in $C^{2,1}$ with the fibration basis being
the two-supersphere in $R^{3,2}$.

It is worth noting that
$\Sigma_i$
and $F_{\mu}$ are  respectively even and odd generators of the $osp(2,1)$
superalgebra:
\[
[\Sigma_3 ,\Sigma_{\pm} ]\ =\ \pm \Sigma_{\pm} \ ,\ [\Sigma_+ ,\Sigma_- ]\ =
\ 2\Sigma_3 \ ,
\]
\[
[\Sigma_3 ,F_{\pm} ]\ =\ \pm \frac{1}{2} F_{\pm} \ ,
[\Sigma_{\pm} ,F_{\pm} ]\ =\ 0\ ,\ [\Sigma_{\pm} ,F_{\mp} ]\ =\ F_{\pm} \ ,
\]
\be
[F_{\pm} ,F_{\pm} ]\ =\ \pm \Sigma_{\pm} \ ,\ [F_+ ,F_- ]\ =\ -\Sigma_3 \ .
\ee
Here and in what follows, the symbol $[A,B]$ denotes a supercommutator, i.e.
the commutator $AB-BA$ if either $A$ or $B$ is even, and the anticommutator
$AB+BA$ if both $A$ and $B$ are odd.

A superfunction $\Phi = \Phi (\xi ,\xi^* )$ on $sS^3$ is represented as a
linear combination of monomials
\be
{\chi }_1^{*m_1} {\chi }_2^{*m_2} {\chi }_1^{n_1} {\chi }_2^{n_2}
a^{*\mu} a^{\nu} \ ,
\ee
where $m_{\alpha}$, $ n_{\alpha}$ are non-negative integers and
$\mu,\nu = 0,1$\footnote{If the monomial in (59) is odd, then it appears
with an odd coefficient in the decomposition of the superfunction $\Phi$
in the linear combination of the monomials (59). Thus the
superfunction $\Phi$  itself is an even element of the
Grassmann algebra.}.
The representatives,  which are identical on the
surface $\chi_1^*\chi_1 +\chi_2^*\chi_2 +a^*a=r$, correspond to the same
superfunction $\Phi$ on $sS^3$.
To any such monomial we assign the topological charge
$2k = m_1 +m_2 +\mu -n_1 -n_2 -\nu$. As $s{\cal H}_k$,
$k \in \frac{1}{2}{\bf Z}$, we denote the linear space spanned by the
monomials with fixed $k$. Any superfunction $\Phi \in s{\cal H}_k$ can
be expanded as
\be
\Phi \ =\ \Phi_0 (\chi ,\chi^* )\ +\ f(\chi ,\chi^* )a\
+\ g(\chi ,\chi^* )a^* \ +\ F(\chi ,\chi^* )a^* a\ ,
\ee
where $\Phi_0 , F \in {\cal H}_k$ are even and
 $f \in {\cal H}_{k+\frac{1}{2}}$,
$g \in {\cal H}_{k-\frac{1}{2}}$ are odd.
 The space
$s{\cal H}_0$ is a superalgebra $s{\cal A}$ with respect to the
supercommutative `point-wise' product of parameters $\xi ,\xi^*$. The spaces
$s{\cal H}_k$ are $s{\cal A}$-bimodules.

The differential operators generating the $osp(2,1)$ algebra acting on
$s{\cal H}_k$ are given by
\[
J_i \ =\ \frac{i}{2} \ [\xi^*_a \Sigma^i_{ab} \partial_{\xi^*_b} \ -\
\xi_a \Sigma^{*i}_{ab} \partial_{\xi_b} ]\ ,
\]
\be
v_{\mu} \ =\ \frac{i}{2} [\xi^*_a F^{\mu}_{ab} \partial_{\xi^*_b} \ +\
\xi_a F^{*\mu}_{ab} \partial_{\xi_b} ]\ .
\ee
The function $C(x,\theta )$ $=$
$x^2_i +\frac{1}{2} (\theta_+ \theta_- - \theta_- \theta_+ )$ is an
invariant superfunction from $s{\cal H}_0$:
\be
J_i C(x,\theta ) \ =\ 0\ ,\ v_{\mu} C(x,\theta ) \ =\ 0\ .
\ee
The condition $C(x,\theta )\ =\ \rho^2$ defines a supersphere in
${\bf R}^{3,2}$.

In $s{\cal H}_0$ we can introduce a standard Berezin integral over the
supersphere \cite{GKP2} as
\be
sI_{\infty} [\Phi ]\ =\ \frac{\rho}{\pi} \int d^3 x d\theta_+ d\theta_-
\delta (x^2_i + \theta_+ \theta_- - \rho^2 )\ \Phi (x,\theta)\ ,
\ee
where the super $\delta$-function should be understood as
\[
\delta (x^2_i + \theta_+ \theta_- -\rho^2 )\ =\ \delta (x^2_i -\rho^2 )
\ +\ \theta_+ \theta_- \ {\delta }' (x^2_i -\rho^2 )\ ,
\]
(this formula is obtained by a Taylor expansion of the super
$\delta$-function). Expressing $\theta_{\mu}$ in terms of $a, a^*$ we
obtain
\be
sI_{\infty} [\Phi ]\ =\ \frac{1}{\pi} \int d^3 x da^* da
[\delta (x^2_i - \rho^2 )\ +\rho a^* a{\delta }'(x^2_i -\rho^2 )]\
\Phi (x,a)\ ,
\ee
where  one can now expand $\Phi$ directly as in (60). Much as before, we
can introduce the inner product in $s{\cal H}_k$
\be
(\Phi_1 ,\Phi_2 )_k \ =\ sI_{\infty} [\Phi^*_1 \Phi_2 ] \ ,
\ee
where (cf. (60))
\[ \Phi^*=\Phi_0^*(\chi,\chi^*) -g^*a +f^*a^* +F^*a^*a .\]
The spinor field $\Psi$ we identify with the odd part of the superfield
$\Phi$:
\be
\Psi \ =\ f(\chi ,\chi^* )a\ +\ g(\chi ,\chi^* )a^* \ ,
\ee
where $f$ and $g$ are also odd (cf. footnote 5).
As ${\cal S}_k$, $k \in \frac{1}{2}{\bf Z}$ we denote the set of spinor
fields from $s{\cal H}_k$, i.e. with
$f \in {\cal H}_{k+\frac{1}{2}}$ and $g \in {\cal H}_{k-\frac{1}{2}}$;
${\cal S}_k$ are $\cal A$-bimodules (but not $s{\cal A}$-(bi)modules).
The formula (66) induces the decomposition of the spinor space:
\be
{\cal S}_k \ =\ {\cal S}^{(+)}_k \ \oplus \ {\cal S}^{(-)}_k \ ,
\ee
where ${\cal S}^{(+)}_k$ corresponds to the first term in (66), and
${\cal S}^{(-)}_k$ to the second one. The chirality grading operator
$\Gamma$ we define as the operator taking in ${\cal S}^{(\mu)}_k$ the
value $\mu = \pm 1$:
\be
\Gamma \Psi \ =\ f(\chi ,\chi^* )a\ -\ g(\chi ,\chi^* )a^* \ .
\ee
The operator $\Gamma$ can be realized as the differential operator
\be
\Gamma \Psi \ =\ (a \partial_a \ -\ a^* \partial_{a*} )\Psi \ .
\ee

We define the free Dirac operator $D$ as the following mapping from
${\cal S}_k$ to ${\cal S}_k$:
\be
D \Psi \ =\ (K_+ g)a\ +\ (K_- f)a^* \ .
\ee
It can be expressed as the second-order (in even and odd parameters)
differential operator
\be
D \Psi \ =\ (a\partial_{a^*} K_- \ +\
a^* \partial _a K_+ )\Psi \ .
\ee
The Dirac operator anticommutes with the chirality operator:
\be
D\Gamma \ +\ \Gamma D\ =\ 0\ .
\ee

The action for spinor field we define in terms of the Berezin integral
\be
S[\Psi ,\Psi^* ]\ =\ sI_{\infty} [\Psi^* D \Psi \ +\ W(\Psi^* ,\Psi )]\ ,
\ee
where $W(\Psi^* ,\Psi )$ is a suitable potential describing the
self-interaction of the spinor field.

\noindent {\it Note}: We would like to stress that the formalism for
 the spinor field
presented above is equivalent to the usual one described in Section 2, but
it is better suited for the non-commutative generalization.

In the noncommutative case, we replace $\chi_{\alpha}$, $\chi^*_{\alpha}$,
and $a$, $a^*$ by bosonic and fermionic annihilation and creation operators
\[
\hat\chi_{\alpha} \ =\ A_{\alpha} R^{-1/2} \ ,\
\hat\chi^*_{\alpha} \ =\  R^{-1/2} A^*_{\alpha} \ ,
\]
\be
\hat{a}\ =\ b R^{-1/2} \ ,\
\hat{a}^* \ =\ R^{-1/2} b^* \ ,
\ee
where now
\be
R = A^*_{\alpha} A_{\alpha} \ +\ b^* b\ ,
\ee
so that the condition $\chi^*_{\alpha} \chi_{\alpha} + a^* a = 1$ is
satisfied (this corresponds to the radius of the two-sphere $\rho = 1/2$).
The operators $A_{\alpha} , A^*_{\alpha} , b, b^*$ act in the Fock space
$s{\cal F}$ spanned by the orthonormal vectors
\[
|n_1 ,n_2 ;\nu \rangle =
\frac{1}{\sqrt{n_1 ! n_2 !}}
(A^*_1)^{n_1} (A^*_2)^{n_2} (b^*)^{\nu} |0 \rangle \ ,
\]
where $n_1 ,n_2$ are non-negative integers, $\nu = 0,1$ and  $|0 \rangle$
is the vacuum defined by
$A_1 |0 \rangle = A_2 |0 \rangle = b |0 \rangle = 0 $\footnote{Note that
* still denotes the {\it graded} involution. This means that $b^*$ is the
adjoint operator $b^{\dagger}$ of $b$ in the standard fermionic Fock
space, but $b^{**}=-b$.}.
The annihilation and creation operators in question satisfy in $s{\cal F}$
the supercommutation relations
\[
[A_{\alpha} , A_{\beta} ]\ =\ [A^*_{\alpha} , A^*_{\beta} ]\ =\ 0\ , \
[A_{\alpha} , A^*_{\beta} ]\ =\ \delta_{\alpha ,\beta} \ .
\]
\[
[A_{\alpha} ,b]\ =\ [A^*_{\alpha} ,b]\ =\
[A_{\alpha} ,b^* ]\ =\ [A^*_{\alpha} ,b^* ]\ =\ 0\ , \
\]
\be
b^2 \ =\ (b^* )^2 \ =\ 0\ ,\ [b,b^* ]\ =\ 1\ .
\ee
The operators
\[
R_j \ =\ \frac{1}{2}\ A^*_{\alpha} \sigma^j_{\alpha \beta} A_{\beta} \ ,
\]
\be
V_+ \ =\ \frac{1}{\sqrt{2}} (-A^*_1 b\ -\ b^* A_2 )\ ,
V_- \ =\ \frac{1}{\sqrt{2}} (-A^*_2 b\ +\ b^* A_1 )\ ,
\ee
then satisfy  in the Fock space $s{\cal F}$ the $osp(2,1)$ superalgebra
commutation relations (58) (with an obvious change in the notation).

As $s{\hat {\cal H}}_k$ we denote the linear space of superfields (60) with
$\Phi_0 ,F \in {\hat {\cal H}}_k$, $f \in {\hat {\cal H}}_{k+\frac{1}{2}}$
and $g \in {\hat {\cal H}}_{k-\frac{1}{2}}$. Obviously,
$s{\cal A} = s{\hat {\cal H}}_0$ is a superalgebra, and $s{\hat {\cal H}}_k$
are $s{\cal A}$-bimodules. The subalgebra ${\cal A}$ is naturally embedded
into $s{\cal A}$ (as the set of ${\Phi_0}$'s in the decomposition (60) for
$k=0$). The generators  $J_j$ and ${\cal V}_{\mu}$ of $osp(2,1)$ act in the
space
$s{\hat {\cal H}}_k$ by means of the superadjoint action (cf. \cite{GKP2})
\be
J_j \Phi \ =\ [R_j ,\Phi ]\ ,\ {\cal V}_{\mu}\Phi \ =\ [V_{\mu} ,\Phi ]\ .
\ee
Obviously, this is a reducible representation of the superalgebra $osp(2,1)$ in
$s{\hat {\cal H}}_k$.

The spinor field $\Psi$ we identify with the odd part of the superfield
$\Phi$:
\be
\Psi \ =\ f(\hat\chi^* ,\hat\chi )\hat{a}\ +\ g(\hat\chi^* ,\hat\chi )
\hat{a}^* \ .
\ee
As ${\hat {\cal S}}_k$, $k \in \frac{1}{2}{\bf Z}$ we denote the set of
spinor fields from $s{\hat {\cal H}}_k$, i.e. with
$f \in {\hat {\cal H}}_{k+\frac{1}{2}}$ and
$g \in {\hat {\cal H}}_{k-\frac{1}{2}}$. ${\hat {\cal S}}_k$ are
$\cal A$-bimodules (but not $s{\cal A}$-(bi)modules).

Formula (79) induces the decomposition of the spinor space:
\be
{\hat {\cal S}}_k \ =\ {\hat {\cal S}}^{(+)}_k \
\oplus \ {\hat {\cal S}}^{(-)}_k \ ,
\ee
where ${\hat {\cal S}}^{(+)}_k$ corresponds to the first term in (79), and
${\hat {\cal S}}^{(-)}_k$ to the second one. The grading $\hat\Gamma$ we define
in the same way as in (68):
\be
\hat\Gamma \Psi =  f(\hat\chi^* ,\hat\chi )\hat{a}\ -\
g(\hat\chi^* ,\hat\chi )\hat{a}^*.
\ee
It can  also be written as
\be
\hat\Gamma \Psi \ =\ -\ [b^* b,\Psi ]\ .
\ee
Thus the grading is directly related to the fermion number and it takes, in
${\hat {\cal S}}^{(\mu)}_k$, the value $\mu = \pm 1$.

The free Dirac operator $D$ we define  as in (70):
\be
D \Psi \ =\  (K_+ g)\hat{a} \ +\
(K_- f)\hat{a}^*  \ ,
\ee
where $K_{\pm}$ were defined in (50).
The Dirac operator maps ${\hat {\cal S}}_k$ into ${\hat {\cal S}}_k$  and it
anticommutes with the chirality grading operator:
\be
D\Gamma \ +\ \Gamma D\ =\ 0\ .
\ee

As $s{\cal F}_N$ we denote the subspace
\be
s{\cal F}_N \ =\ \{ |n_1 ,n_2 ;\nu \rangle \in s{\cal F} \ ,\
n_1 +n_2 +\nu = N \} \ ,
\ee
in which the operator $R$ takes the value $R=N$. Obviously, $s{\cal F}_N$
is a $(2N+1)$-dimensional superspace
\be
s{\cal F}_N \ =\ s{\cal F}^{(0)}_N \ \oplus \ s{\cal F}^{(1)}_{N-1} \ ,
\ee
where $s{\cal F}^{(\nu)}_{N'}$ is the subspace with $N'$ bosons and $\nu$
fermions, $s{\cal F}^{(0)}_N$ is the even subspace of $s{\cal F}_N$, and
$s{\cal F}^{(1)}_{N-1}$ is the odd one. The space $s{\cal F}_N$ is the
representation space of the irreducible representation of the superalgebra
$osp(2,1)$ (generated by $R_j ,V_{\mu}$ given above) in which the Casimir
operator
\be
C = R^2_3 \ +\ \frac{1}{2} (R_+ R_- \ +\ R_- R_+ )\ +\
\frac{1}{2} (V_+ V_- - V_- V_+ )
\ee
takes the value
\be
C = \frac{1}{4} N(N + 1)\ .
\ee

As $s{\hat {\cal H}}_{MN}$ we denote the space of linear mappings from
$s{\cal F}_N$ to $s{\cal F}_M$ spanned by the operator monomials
\[
\hat\chi^{*m_1}_1 \hat\chi^{*m_2}_2 \hat\chi^{n_1}_1 \hat\chi^{n_2}_2
\hat{a}^{*\mu} \hat{a}^{\nu}
\]
with $m_1 +m_2 +\mu \leq M$, $n_1 +n_2 +\nu \leq N$ and
$m_1 +m_2 +\mu -n_1 -n_2 -\nu = M-N$. Any operator
$\Phi \in s{\hat {\cal H}}_{MN}$ can be represented by a
$(2N+1) \times (2M+1)$ supermatrix. In $s{\hat {\cal H}}_{MN}$ we introduce
an inner product
\be
(\Phi_1 ,\Phi_2 )_{MN} \ =\ sTr_N (\Phi^*_1 \Phi_2 )\ ,
\ee
where $J=\frac{1}{2}(M+N)$ and $sTr_N$ denotes the supertrace in the
space $s\hat{\cal H}_{NN}$. As in the purely bosonic case, the action of
$su(2)$ generators $J_i$ on $s\hat{\cal H}_{MN}$ is given by the formula (44)
with $R_i$ given by (77).

As ${\hat {\cal S}}_{MN}$ we denote the space of spinor fields from
$s{\hat {\cal H}}_{MN}$, i.e.
\[
\Psi \ =\ f(\hat\chi^* ,\hat\chi )\hat{a}\ +\ g(\hat\chi^* ,\hat\chi )
\hat{a}^* \ ,
\]
where $f \in {\hat {\cal H}}_{M,N-1}$ and
$g \in {\hat {\cal H}}_{M-1,N}$. This gives the decomposition
\[
{\hat {\cal S}}_{MN} \ =\ {\hat {\cal S}}^{(+)}_{MN} \
\oplus \ {\hat {\cal S}}^{(-)}_{MN} \ ,
\]
where ${\hat {\cal S}}^{(\mu)}_{MN}$ contains spinor fields with the
chirality $\mu =\pm 1$. The operators $K_{\pm}$, entering the Dirac operator
$D$ (83), act in
${\hat {\cal S}}_{MN}$; moreover, $K_0$ takes in ${\hat {\cal S}}^{(\mu)}_{MN}$
the constant value $\frac{1}{2}(M-N+\mu )$. The spinor field operators
from ${\hat {\cal S}}^{(\mu)}_{MN}$ are odd mappings
$s{\cal F}_N \to s{\cal F}_M$, namely:
\be
f\hat{a}:\ s{\cal F}^{(1)}_N \rightarrow s{\cal F}^{(0)}_M \ ,
\qquad
g\hat{a}^* :\ s{\cal F}^{(0)}_N \rightarrow s{\cal F}^{(1)}_M \ .
\ee
According to (45), $f$ and $g$ can be expanded into operator functions
belonging to the representations:
\[
\frac{M}{2} \otimes \frac{N-1}{2} \ =\ |k+\frac{1}{2}| \ \oplus \ ...\
\oplus \ (J-\frac{1}{2} )\ ,\ {\rm for}\ f\ ,
\]
\be
\frac{M-1}{2} \otimes \frac{N}{2} \ =\ |k-\frac{1}{2}| \ \oplus \ ...\
\oplus \ (J-\frac{1}{2} )\ ,\ {\rm for}\ g\ ,
\ee
where $k=\frac{1}{2}(M-N)$ and $J=\frac{1}{2}(M+N)$. This means that $f$ can
be expanded into the functions $\Phi^{j}_{J-\frac{1}{2},k+\frac{1}{2},m}$ with
$j=|k+\frac{1}{2}|, \dots ,J-\frac{1}{2}$, $|m|\leq j$, and, in the same way,
$g$ into the functions $\Phi^{j}_{J-\frac{1}{2},k-\frac{1}{2},m}$ with
$j=|k-\frac{1}{2}|, \dots ,J-\frac{1}{2}$, $|m|\leq j$ (cf. (48)).
The admissible
values of $j$ are:
\[
j\ =\ |k|-\frac{1}{2} \ ,\ |k|+\frac{1}{2},\ \dots \ ,\
J-\frac{1}{2} \ .
\]
They can be seen from the representation content in Eqs. (91):

(i) The first value $j=|k|-\frac{1}{2}$ is admissible only for $k\neq 0$,
and for $k>0$ it is present in $g$ (chirality $-1$), whereas for $k<0$
it is present in $f$
(chirality $+1$). It corresponds to the $|M-N|$ (normalized)
zero modes of the Dirac
operator given by
\[
\Psi^{0+}_{m_1 m_2} \ =\ \frac{\sqrt{(m_1+m_2+1)!}}{\sqrt{m_1 ! m_2 !}} \
\hat\chi_1^{*m_1} \hat\chi_2^{*m_2} \hat{a}^* \ ,\
{\rm for}\ k=\frac{1}{2} (m_1 +m_2+1 ) > 0\ ,
\]
\be
\Psi^0_{n_1 n_2} \ =\ \frac{\sqrt{(n_1+n_2+1)!}}{\sqrt{n_1 ! n_2 !}} \
\hat\chi_1^{n_1} \hat\chi_2^{n_2} \hat{a} \ ,\ {\rm for}\
k=-\frac{1}{2} (n_1 +n_2 +1) < 0\ ,
\ee
Note that the normalization does not depend on the cut-off spin $J$; therefore
the correct commutative limit is obvious.

(ii) The remaining eigenvalues $j=|k|+\frac{1}{2},\ \dots ,\ J-\frac{1}{2}$,
correspond to non-zero modes of the Dirac operator. Consider  (normalized)
functions
$\Phi^{j\pm}_{Jkm}$ with a given chirality $\mu =\pm 1$
 given by
\be
\Phi^{j-}_{Jkm} \ =
\ \Phi^{j}_{J-\frac{1}{2},k-\frac{1}{2},m}\ \hat{a}^* \ ,\
\Phi^{j+}_{Jkm} \ =
\ \Phi^{j}_{J-\frac{1}{2},k+\frac{1}{2},m}\ \hat{a} \ ,
\ee
where $\Phi^{j}_{Jk'm}$ were defined in (46) and (48).
 Using the definition of the Dirac
operator (83) and the formula\footnote{Formula (94) follows from the properties
of the Wigner functions (cf. the discussion after (48)) because $K_{\pm}$
act on the second subscript of $\Phi^j_{Jlm}$ in exactly the same way as
$J_{\pm}$ act on the third subscript.}
\be
K_{\pm} \Phi^{j}_{Jlm}\ =\ \sqrt{(j\pm l+1)(j\mp l)} \
\Phi^{j}_{J,l\pm 1,m} \ ,
\ee
we obtain the equation
\be
D\Phi^{j\pm}_{Jkm} \ =\
E_{kj} \Phi^{j\mp}_{Jkm} \ ,
\ee
where
\be
E_{kj} \ =\
\sqrt{\left( j+\frac{1}{2} \right)^2  - k^2} \ .
\ee
 Thus the functions
\be
\Psi^{j\pm}_{Jkm} \ =\frac{1}{\sqrt{2}} \ \Phi^{j+}_{Jkm}\ \pm
\  \frac{1}{\sqrt{2}}\ \Phi^{j-}_{Jkm}
\ee
are (normalized)
eigenfuctions of the Dirac operator with the eigenvalues $\pm E_{kj}$.

\noindent {\it Note}: We would like to stress that in the standard
commutative case
exactly the same formula is obtained for the spectrum, except for
the fact that
the  admissible values of $j$ are not truncated. As the number of degrees
of freedom is finite this will lead to a non-perturbative UV-regularization
(for a real scalar field this was demonstrated in \cite{GKP1}).

The action for the self-interacting (Dirac) spinor field
$\Psi \in {\hat {\cal S}}_{MN}$ with a fixed winding number
we define as follows
\be
S_{MN} [\Psi^* ,\Psi ]\ =\ \ sTr_N
[\Psi^* D \Psi \ +\ W(\Psi^*, \Psi )]\ ,
\ee
where $\Psi = f a + g a^*$, $\Psi^* = ag^*- a^*f^*$, and $W(.,.)$ is
an interaction term.
We do not wish to fix  the winding number of the field $\Psi$;
instead we take $\Psi$ from
the space
\be
{\hat{\cal S}}_{(J)} \ =\ \bigoplus_{M+N=2J} {\hat{\cal S}}_{MN} \ ,
\ee
and we define the corresponding action  as
\be
S_{(J)} [\Psi^* ,\Psi ]\ =\ \sum_{M+N=2J} S_{MN} [\Psi^* ,\Psi ]\ .
\ee
This action has the following basic properties:

1) It is invariant with respect to the space isometries, i.e. the rotations
of the sphere, and the chiral transformations
\be
\Psi \to e^{i\alpha \Gamma} \Psi \ ,
\Psi^* \to \Psi^* e^{i\alpha \Gamma}  \ , \alpha \in {\bf R} \ ,
\ee
provided that the interaction term is rotationally and chirally invariant.

2) It describes the system with a finite number of degrees of freedom.

3) It approaches for $J \to \infty$ the commutative action.

\noindent {\it Note}: The rotational invariance of the action $S_{MN}$ follows
from the rotational invariance of the truncated Dirac operator.
The chiral invariance of the action is obvious, because the Dirac operator $D$
anticommutes with the grading $\Gamma$ (cf. (84)).

The spinor field $\Psi \in {\hat {\cal S}}_{MN}$ can be expanded as
\be
\Psi \ =\ \sum_{m_1 m_2} a^{0,|k|-\frac{1}{2}}_{m_1 m_2} \
\Psi^0_{m_1 m_2} \ +\ \sum_{j=|k|+\frac{1}{2}}^{J-\frac{1}{2}}
\sum_{m=-j}^{+j} \ \left( a^{j+}_{km} \ \Phi^{j+}_{Jkm} \ + \
a^{j-}_{km} \ \Phi^{j-}_{Jkm} \right) \ ,
\ee
where the first sum corresponds to the zero modes (92) and the remaining two
sums are related to the non-zero modes (96). All expansion coefficients
$a^{..}_{..}$ in (102) are supposed to be independent anticommuting
Grassmannian variables. In the same way,
 the field $\Psi^* \in {\hat {\cal S}}_{NM}$
is supposed to have an expansion with independent Grassmann coefficients
$a^{*..}_{..}$.

The quantum field mean value of a functional $F[\Psi ,\Psi^* ]$
is given as
\be
\langle F[\Psi ,\Psi^* ] \rangle_{MN} \ =\ \frac{\int [D\Psi]_{MN}
[D\Psi^*]_{NM}
e^{-S_{MN} [\Psi ,\Psi^* ]} F[\Psi ,\Psi^* ]}{\int [D\Psi]_{MN} [D\Psi^* ]_{NM}
e^{-S_{MN} [\Psi ,\Psi^* ]}} \ ,
\ee
where $\int [D\Psi]_{MN} [D\Psi^* ]_{NM} \dots$ denotes the finite-dimensional
Berezin integral over all admissible coefficients $a^{..}_{..}$ and
$a^{*..}_{..}$ with fixed $k=\frac{1}{2}(M-N)$ and $J=\frac{1}{2}(M+N)$.

Taking into account that $\Phi^{j+}_{Jkm}$ and the zero modes for $k>0$
correspond to $g$, and similarly $\Phi^{j-}_{Jkm}$ and the zero modes for
$k<0$ correspond to $f$, the chiral transformations (101) can be written as
\be
a^{0,|k|-\frac{1}{2}}_{m_1 m_2} \to \ e^{-i\varepsilon (k)\alpha} \
a^{0,|k|-\frac{1}{2}}_{m_1 m_2} \ ,
a^{j\pm }_{km} \to e^{\mp i\alpha} a^{j\pm }_{km} \ ,
\ee
where $\varepsilon (k)$ is the sign function. We see that the phase
contributions to the measure $[D\Psi]_{MN}$ from the non-zero modes cancel.
As expected, the only contribution comes from $|\kappa |$ zero modes
($\kappa = M-N$):
\[
[D\Psi ]_{MN} \to e^{-i\kappa \alpha} \ [D\Psi ]_{MN} \ .
\]
Analogous rules are valid for expansion coefficients of the field $\Psi^*$.
The phase factors from non-zero modes cancel, and there is just the same
phase contribution from zero modes (since the conjugation and the change
$M-N \to N-M$ compensate each other in the phase factors):
\[
[D\Psi^* ]_{NM} \to e^{-i\kappa \alpha} \ [D\Psi^* ]_{NM} \ .
\]
Thus, the total change of the measure $[D\Psi ]_{MN} [D\Psi^* ]_{NM}$ is
\be
[D\Psi ]_{MN} [D\Psi^* ]_{NM} \
\to \ e^{-2i\kappa \alpha} \ [D\Psi ]_{MN} [D\Psi^* ]_{NM} \ .
\ee

We see that assuming fixed $k \neq 0$, the chiral symmetry is violated on
quantum level. However,
 simultaneously taking into account fields with given $k$
and $-k$, as e.g. in ${\hat{\cal S}}_{(J)}$, the chiral symmetry is restored.
Thus, the measure $D\Psi D\Psi^* \ =\
\prod_{MN} [D\Psi ]_{MN} [D\Psi^* ]_{NM}$ entering the quantum mean value
over ${\hat{\cal S}}_{(J)}$,
\be
\langle F[\Psi ,\Psi^* ] \rangle \ =\ \frac{\int D\Psi D\Psi^*
e^{-S_{(J)} [\Psi ,\Psi^* ]} F[\Psi ,\Psi^* ]}{\int D\Psi D\Psi^*
e^{-S_{(J)} [\Psi ,\Psi^* ]}} \ ,
\ee
is invariant under chiral transformations (101) or (104),
 since  $D\Psi D\Psi^*$
$\to$ $D\Psi D\Psi^*$.

\section{Summary and outlook}
In treating  the topologically nontrivial complex scalar field
configurations in the noncommutative case,
our main tool was the noncommutative version of
the Hopf fibration encoded in the noncommuting parameters $\hat\chi_{\alpha}$,
$\hat\chi^*_{\alpha}$, $\alpha = 1,2$. Any field configuration with the
topological winding number $\kappa$ was expanded into the functions
$\Phi^{j}_{Jkm}$, $k=\frac{1}{2} (M-N)$,
$|m|\leq j=|k|, \dots ,\frac{1}{2}(M+N)$ where $\Phi^{j}_{Jkm}$ are
noncommutative analogues of the standard $D$-functions $D^j_{km}$.
 Thus we gave
an algebraic characterization of the winding number $\kappa =2k$, which is
directly related to the index $k$ of the $D$-functions in question. On the
matrix level this leads to the $(M+1)\times (N+1)$-matrix representation of
fields from the space ${\hat {\cal H}}_{MN}$. The usual matrix geometry models
correspond to $M=N$, and this is the reason why they describe the
topologically trivial configurations only.

The same procedure applied  for the treatment of the
topologically nontrivial spinor field
configurations, too. Moreover, here
 the natural supersymmetry
of the problem introduced in \cite{GKP2} was essential;
we described the topologically nontrivial
spinors on $S^2$ as the odd sections of nontrivial super-line bundles on the
supersphere $sS^2$.
 For this purpose we used the noncommutative
Hopf superfibration in terms of even noncommutative parameters
$\hat\chi_{\alpha}$, $\hat\chi^*_{\alpha}$, $\alpha = 1,2$, to which we
added a pair
$\hat{a}, \hat{a}^*$ of odd noncommutative parameters. We identified the
spinor bundle
${\hat {\cal S}}_{MN}$ as the smallest space that is invariant under action
of the Dirac operator $D$. The bundle ${\hat {\cal S}}_{MN}$ is spanned by
$|M-N|$ zero modes of $D$ and by the functions
$\Phi^{j}_{J-\frac{1}{2},k-\frac{1}{2},m} \hat{a}^*$,
$\Phi^{j}_{J-\frac{1}{2},k+\frac{1}{2},m} \hat{a}$ corresponding to the
 non-zero
modes of $D$.

The models (98) and (100)  are rotationally invariant and contain only a finite
number of degrees of freedom on both  classical and quantum levels. This
truncation of the modes has the  consequence that their quantum version is
UV-regular (there are only finite sums instead of singular integrals and/or
infinite series).  A detailed discussion
of these aspects in the case
of a real scalar field can be found in \cite{GKP1}.

The supersymmetry approach proposed in \cite{GKP2} proved to  be useful in
describing
the chiral properties of spinors. Our spinor-field models have  chirally
invariant actions; however, the field functional measure
$[D\Psi, D\Psi^* ]_{MN}$ is not invariant for a fixed $M-N\neq 0$ and, under
chiral transformations, it is modified due to the zero modes by the factor
$e^{i\kappa \alpha}$. Only when the fields with given $\kappa$ and $-\kappa$
are treated simultaneously is the chiral invariance  recovered. Thus the
chiral
properties of the theory are the same as those in the standard untruncated
case.

It would be desirable to include the gauge fields into our approach. Even
more desirable and important is to extend our scheme to the four-dimensional
(super)sphere $S^4$. We hope to attack these problems in a near
future. \\[10pt]

\noindent {\bf Acknowledgements}
{}~We are grateful to A. Connes,
V. \v{C}ern\'{y}, T. Damour, M. Fecko, J. Fr\"{o}hlich, J. Ft\'{a}\v{c}nik,
K. Gaw\c{e}dzki, J. Madore, H. Neuberger
 and D. Sullivan for useful discussions.

\end{document}